\documentclass[pdftex]{PoS} 

\usepackage{amsmath}
\usepackage{graphicx}
\DeclareGraphicsExtensions{.pdf,.png}

\newcommand{\sgn}{\mathop{\rm sgn}\nolimits}
\newcommand{\dcp}{\delta_\textsc{cp}}
\newcommand{\Dmq}{\Delta m^2}
\newcommand{\eVq}{\ensuremath{\text{eV}^2}}
\newcommand{\Ie}{\textit{i.e.}}
\newcommand{\LT}{\left}
\newcommand{\RT}{\right}

\newenvironment{myitemize} {
  \begin{list}{--}
      {
	\setlength{\leftmargin} {4mm}
	\setlength{\parsep}     {0pt}
	\setlength{\itemsep}    {0mm}
	\setlength{\topsep}     {\itemsep}
	\setlength{\partopsep}  {0pt}
	}} {
  \end{list}}

\title{Prospects and synergies between future atmospheric and
  long-baseline experiments}

\ShortTitle{Prospects and synergies between future atmospheric and
  long-baseline experiments}

\author{\speaker{Michele Maltoni}\\
  Departamento de F\'isica Te\'orica \& Instituto de F\'isica
  Te\'orica UAM/CSIC, Facultad de Ciencias C-XI, Universidad
  Aut\'onoma de Madrid, Cantoblanco, E-28049 Madrid, Spain\\
  E-mail: \email{michele.maltoni@uam.es}}

\abstract{In this talk we will discuss the physics reach of the
  atmospheric neutrino data collected by a future megaton-class
  neutrino detector. First we will discuss the potentialities of
  atmospheric neutrinos on general basis, presenting our results in
  the form of neutrino oscillograms of the Earth: contour plots in the
  neutrino energy--nadir angle plane. In this context we will analyze
  in detail the various signatures related to $\theta_{13}$, to the
  neutrino mass hierarchy, to the octant of $\theta_{23}$ and to the
  $\dcp$ phase which appear for different values of the neutrino
  energy and baseline. Then we will consider concrete experimental
  setups, showing that synergic effects exist between atmospheric and
  long-baseline neutrino data: the combination of the two data sets is
  much more powerful than the simple sum of the sensitivity of each
  individual data sample.}

\FullConference{10th International Workshop on Neutrino Factories,
  Super beams and Beta beams\\
  June 30 - July 5 2008\\
  Valencia, Spain}

\begin{document}

\section{Introduction}
\label{sec:introduction}

Despite their pioneering contribution to the discovery of neutrino
oscillations, it is in general assumed that atmospheric neutrino data
will no longer play an active role in neutrino physics in the coming
years. This is mainly due to the large theoretical uncertainties
arising from the poor knowledge of the atmospheric neutrino fluxes,
which strongly contrast with the requirement of ``precision'' needed
to further enhance our knowledge of the neutrino mass matrix. In this
talk, we will show that despite these large uncertainties the
atmospheric neutrino data collected by a megaton-class detector will
still provide very useful information.

The main strength of atmospheric data is its very broad interval in
neutrino energy ($E_\nu$) and baseline (determined by the nadir angle
$\Theta_\nu$). In order to provide a global view of this whole range,
in this section we will make extensive use of \emph{neutrino
oscillograms} of the Earth, \Ie\ contours plots in the neutrino
energy--nadir angle plane~\cite{Akhmedov:2006hb, Akhmedov:2008qt}.
Consider a bin centered at $(\Theta_\nu,\, E_\nu)$ with extensions
$\Delta\Theta_\nu$ and $\Delta E_\nu$. The numbers of expected
($N_\text{th}$) and observed ($N_\text{ex}$) events in such bin are of
course proportional to its size. Hence its contribution $\Delta\chi^2$
to the total $\chi^2$ function is:
\begin{gather}
    N_\text{th} \simeq \rho_\text{th}(\Theta_\nu,\, E_\nu) \, \Delta S \,, \qquad
    N_\text{ex} \simeq \rho_\text{ex}(\Theta_\nu,\, E_\nu) \, \Delta S \,, \qquad
    \Delta S \equiv \Delta\Theta_\nu \cdot \Delta E_\nu \,;
    \\
    \Delta\chi^2 = 2[N_\text{th} - N_\text{ex}
    + N_\text{ex} \ln(N_\text{ex} / N_\text{th})]
    = [\rho_\text{th} - \rho_\text{ex} 
    + \rho_\text{ex} \ln(\rho_\text{ex} / \rho_\text{th})] \,
    \Delta S \,;
    \\
    \xi^2(\Theta_\nu,\, E_\nu) \equiv
    \lim_{\Delta S \to 0} \frac{\Delta\chi^2}{\Delta S} \,, \qquad
    \xi \equiv \sgn(\rho_\text{ex} - \rho_\text{th}) \sqrt{\xi^2} \,.
\end{gather}
As we will see, different regions of the ($E_\nu$, $\Theta_\nu$) plane
will exhibit characteristic structures whose position and size is
determined by various neutrino parameters. The function $\xi$ provides
an easy way to highlight which of these regions mostly contribute to
the $\chi^2$. Hence in the following we will present isocontours of
$\xi$.

\section{Sensitivity to oscillation parameters}
\label{sec:discussion}

It is well known that for non-zero value of $\theta_{13}$ matter
effects induce a resonance in the $\nu_\mu \to \nu_e$ conversion
probability at $E_\nu \sim 3\div 6$~GeV. The precise position of this
peak in the $(E_\nu,\, \Theta_\nu)$ plane depends on the value of
$\theta_{13}$, so that in principle atmospheric neutrinos could be
used to measure this angle as long as it is larger than about
$3^\circ$. However, in practice the sensitivity is limited since the
$\nu_\mu \to \nu_e$ signal is unavoidably diluted by the $\nu_e \to
\nu_e$ background, and also since the atmospheric neutrino flux at
$E_\nu \sim 6$~GeV is considerably suppressed.
Therefore, although some sensitivity is to be expected in a megaton
detector, it is likely that atmospheric neutrinos will not be
competitive with dedicated long-baseline and reactor experiments in
the determination of $\theta_{13}$.

\begin{figure}[t] \centering 
    \includegraphics[width=0.9\textwidth]{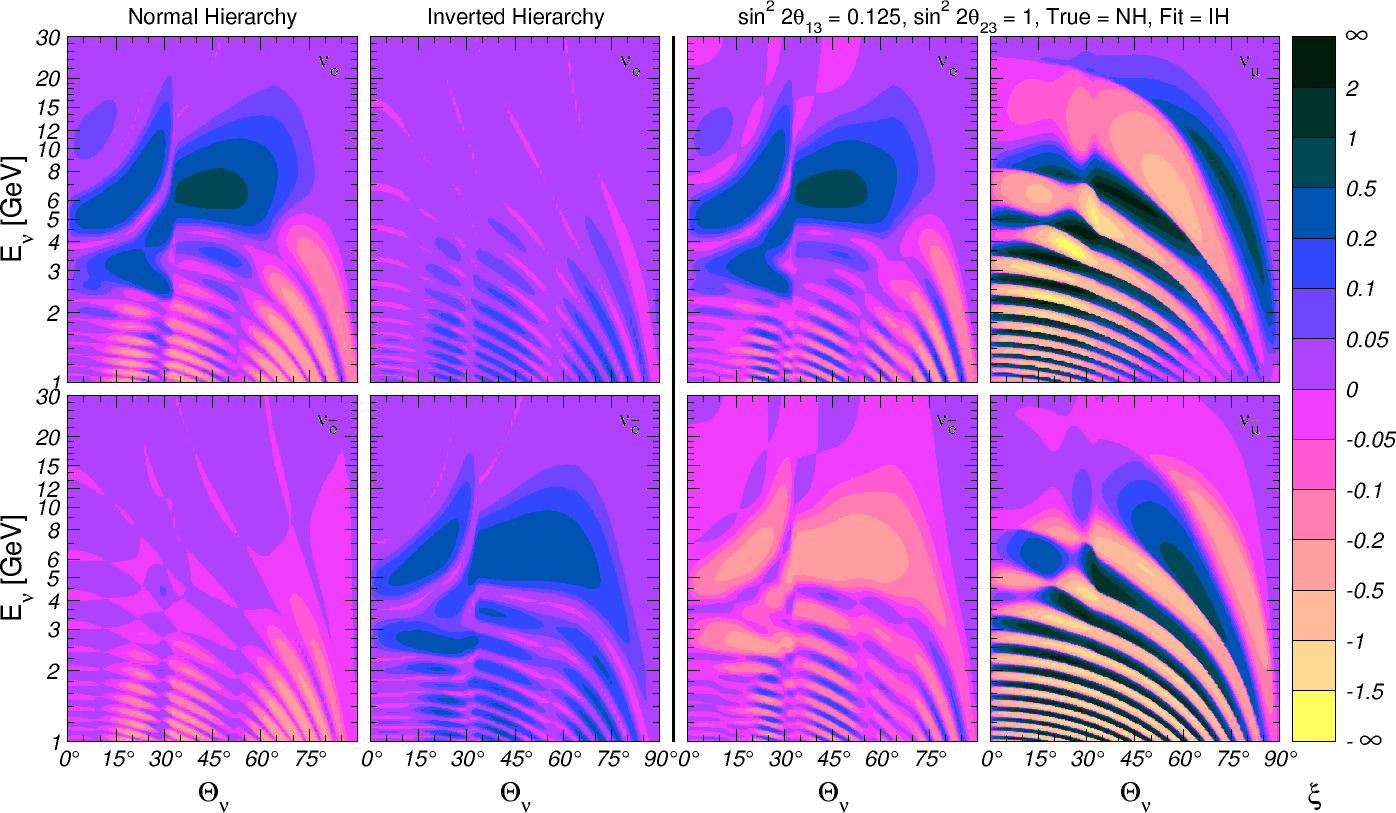}
    \caption{\label{fig:hierar}%
      Sensitivity to the neutrino mass hierarchy. Left: $\sin^2
      2\theta_{13} = 0.125$ ($N_\text{ex}$) versus $\theta_{13} = 0$
      ($N_\text{th}$). Right: normal hierarchy ($N_\text{ex}$) versus
      inverted hierarchy ($N_\text{th}$). The undisplayed parameters
      are $\Dmq_{21} = 8 \times 10^{-5}~\eVq$, $\Dmq_{31} = 2.5 \times
      10^{-3}~\eVq$, $\tan^2\theta_{12} = 0.45$, $\sin^2\theta_{23} =
      1$ and $\dcp = 0$.}
\end{figure}

The sensitivity to the hierarchy is illustrated in
Fig.~\ref{fig:hierar}. We can distinguish two effects:
\begin{myitemize}
  \item in the $\nu_e$ channel there is a peak arising from the
    high-energy resonance just discussed for $\theta_{13}$. This
    signature is only visible if $\theta_{13}$ is large enough. Note
    that in order to determine the hierarchy it is not sufficient to
    see the resonance: it is also necessary to tell whether it
    occurred for neutrinos (normal hierarchy) or antineutrinos
    (inverted hierarchy). It is therefore crucial to have a detector
    capable of charge discrimination. In the case of a non-magnetic
    detector such as a Water Cerenkov, a statistical separation of
    $\nu$ and $\bar\nu$ events can still be obtained by measuring both
    single-ring and multi-ring events, since the relative contribution
    of neutrinos and antineutrinos to these two data sets is
    different.
    
  \item in the $\nu_\mu$ channel we observe a characteristic structure
    of peaks and deeps, which in principle can be used to obtain
    information on the mass hierarchy. However, due to the fast
    alternation of positive and negative regions a very good
    resolution on both neutrino energy and direction is needed,
    otherwise the signal averages to zero and all information is lost.
    Furthermore, the position of the structures is affected also by
    other oscillation parameters, so that a poor knowledge of these
    parameters will result in a considerable loss of sensitivity. Note
    that this signature is also present for $\theta_{13} = 0$, however
    it is strongly suppressed and hardly detectable.
\end{myitemize}
In brief, atmospheric neutrino data can provide useful information on
the neutrino mass hierarchy, but only if $\theta_{13}$ is sufficiently
large and a detector with good resolution and charge discrimination
capability is available.

\begin{figure}[t] \centering 
    \includegraphics[width=0.9\textwidth]{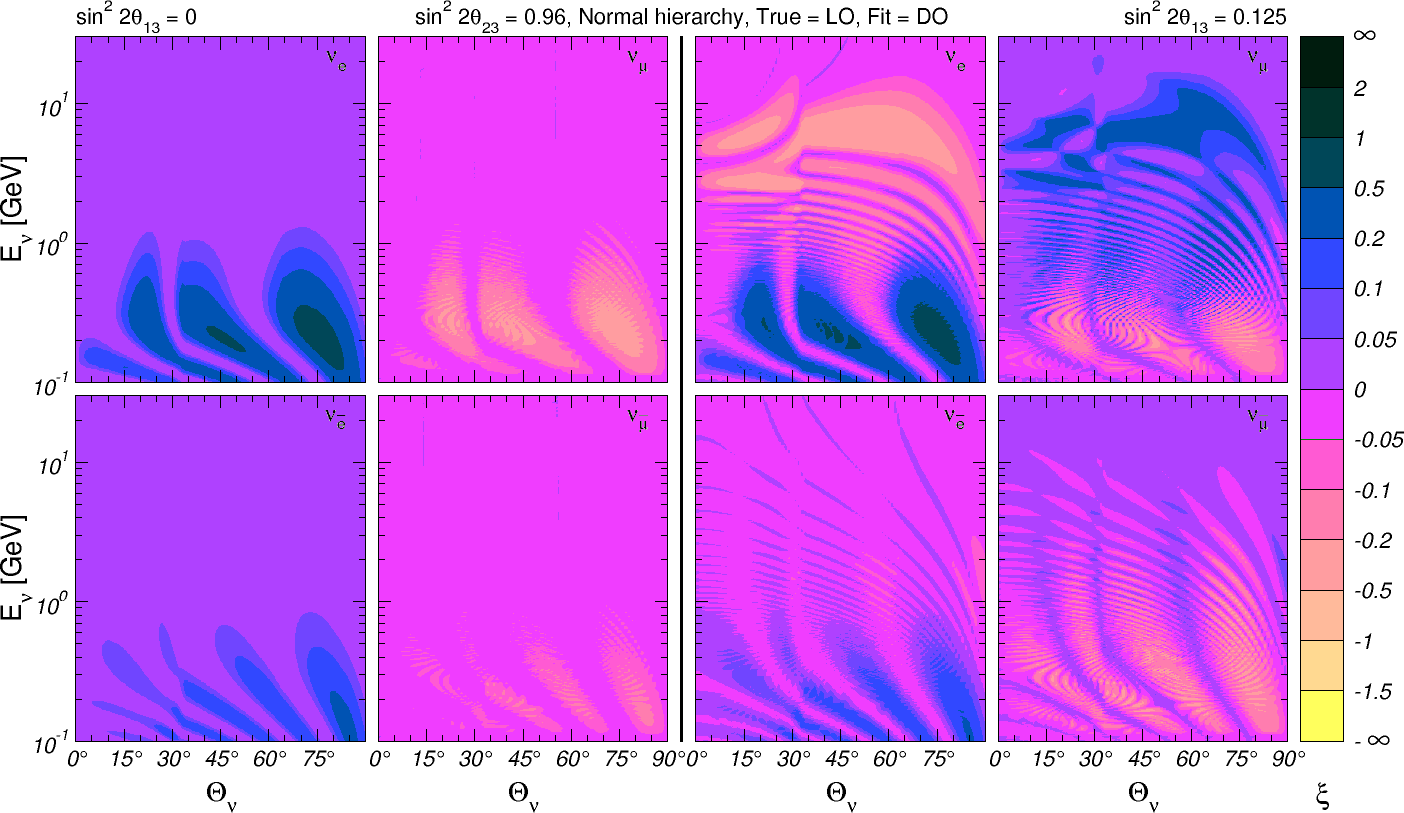}
    \caption{\label{fig:octant}%
      Sensitivity to the octant, for $\theta_{13} = 0$ (left) and
      $\sin^2 2\theta_{13} = 0.125$ (right). Both plots show
      $\sin^2\theta_{23} = 0.4$ ($N_\text{ex}$) versus
      $\sin^2\theta_{23} = 0.6$ ($N_\text{th}$).  The undisplayed
      parameters are $\Dmq_{21} = 8 \times 10^{-5}~\eVq$, $\Dmq_{31} =
      2.5 \times 10^{-3}~\eVq$, $\tan^2\theta_{12} = 0.45$ and $\dcp =
      0$.}
\end{figure}

The sensitivity to the octant is one of the topics where atmospheric
neutrino are mostly useful. As can be seen in Fig.~\ref{fig:octant},
we have two characteristic signatures:
\begin{myitemize}
  \item at \emph{low energy} ($E_\nu < 1$~GeV) we observe an excess of
    $e$-like events and a deficit of $\mu$-like events for
    $\theta_{23} < 45^\circ$ with respect to $\theta_{23} > 45^\circ$.
    The $\nu_e$ signal is about four times larger than the $\nu_\mu$
    one~\cite{GonzalezGarcia:2004cu}. This effect is due to subleading
    oscillations induced by $\Dmq_{21}$, and is present also for
    $\theta_{13} = 0$. For $\theta_{13} \ne 0$ the neutrino flux
    arriving at the detector is modulated with the very fast
    $\Dmq_{31}$ oscillations, but the effect persists on average.
    Since this signature appears with the same sign for neutrinos and
    antineutrinos, no charge discrimination is required for its
    identification.
    
  \item at \emph{high energy} ($E_\nu > 3$~GeV) we observe a deficit
    of $e$-like events and an excess of $\mu$-like events for
    $\theta_{23} < 45^\circ$ with respect to $\theta_{23} > 45^\circ$.
    This effect is again related to the matter resonance discussed for
    $\theta_{13}$ and the hierarchy, and indeed it appears only for
    $\theta_{13} \ne 0$ and only for $\nu$'s or $\bar\nu$'s.
    Therefore, in a detector without charge discrimination this signal
    will be suppressed.
\end{myitemize}
Note that the presence of a low-energy effect independent of
$\theta_{13}$ \emph{guarantees} a minimum sensitivity to the octant
from atmospheric neutrinos, provided that the deviation of
$\theta_{23}$ from maximal mixing is large enough. This is a unique
feature which will prove very synergic with long-baseline data, as we
will show in the next section.

\section{Synergies with long-baseline experiments}
\label{sec:results}

So far we have discussed the potentialities of atmospheric neutrinos
in general terms. Let us now consider three concrete experimental
setups~\cite{Campagne:2006yx}:
\begin{myitemize}
  \item a \emph{Beta Beam} ($\beta$B) from CERN to Fr\'ejus (130~Km).
    We assume 5 years of $\nu_e$ from $^{18}$Ne and 5 years of
    $\bar\nu_e$ from $^6$He at $\gamma = 100$, with an average energy
    $\LT< E_\nu \RT> = 400$~MeV. For the detector we assume the
    MEMPHYS Water-Cerenkov proposal, corresponding to 3 tanks of
    145~Kton each;
    
  \item a \emph{Super Beam} (SPL) from CERN to Fr\'ejus (130~Km).
    We assume 2 years of $\nu_\mu$ and 8 years of $\bar\nu_\mu$
    running, with an average energy $\LT< E_\nu \RT> = 300$~MeV. Again
    we use MEMPHYS as detector;
    
  \item the \emph{T2K phase II} (T2HK) experiment, corresponding to a
    4MW super beam from Tokai to Kamioka (295~Km), with 2 years of
    $\nu_\mu$ and 8 years of $\bar\nu_\mu$. The detector is the
    proposed Hyper-Kamiokande, rescaled to 440~Kton for a fair
    comparison with the $\beta$B and the SPL.
\end{myitemize}

\begin{figure}[t] \centering 
    \includegraphics[width=0.8\textwidth]{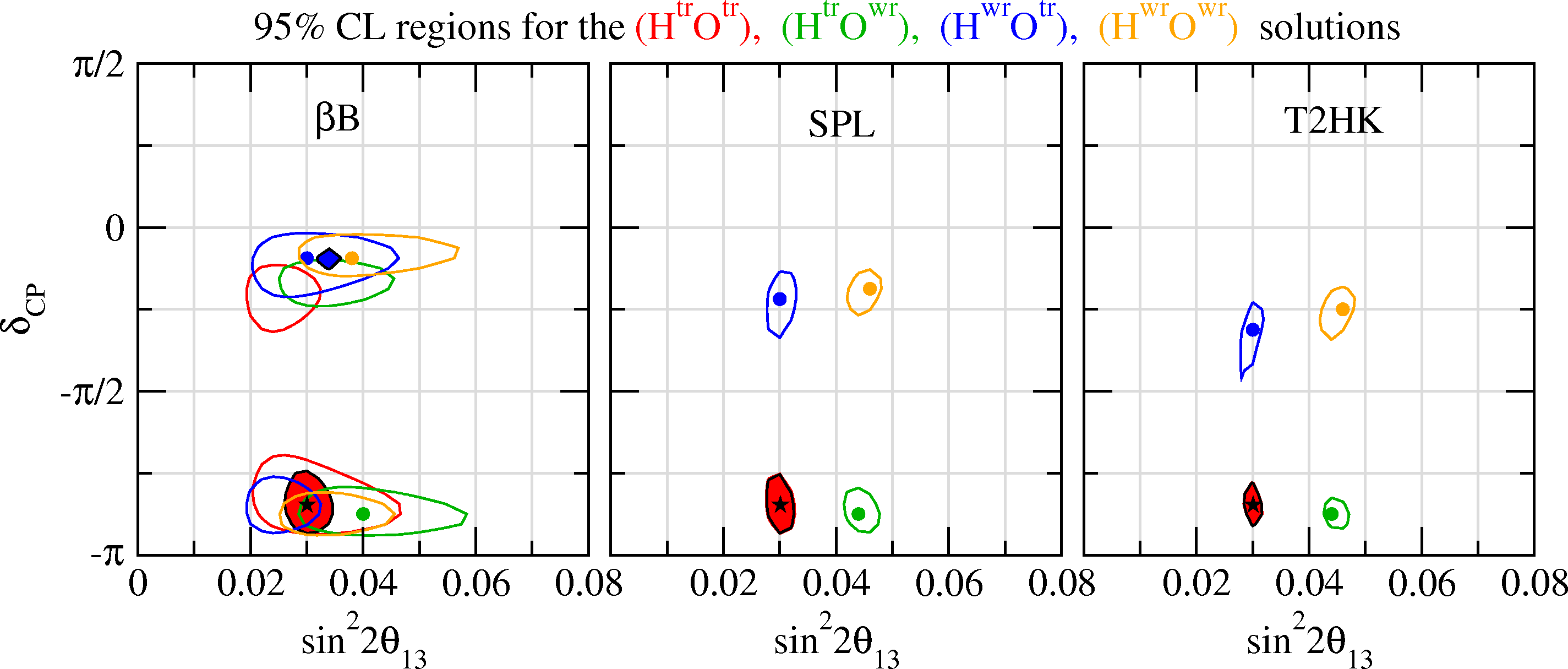}
    \caption{\label{fig:degeneracy}%
      Allowed regions in $\sin^22\theta_{13}$ and $\delta_\mathrm{CP}$
      for LBL data alone (contour lines) and LBL+ATM data combined
      (colored regions)~\cite{Campagne:2006yx}.
      $\text{H}^\text{tr/wr}$ and $\text{O}^\text{tr/wr}$ refers to
      solutions with the true/wrong mass hierarchy and octant,
      respectively. The true parameter values are $\Dmq_{21} =
      7.9\times 10^{-5}~\eVq$, $\Dmq_{31} = 2.4\times 10^{-3}~\eVq$, 
      $\sin^2\theta_{12} = 0.3$, $\sin^22\theta_{13} = 0.03$,
      $\sin^2\theta_{23} = 0.6$ and $\dcp = -0.85 \pi$.}
\end{figure}

As widely discussed at this conference, a characteristic feature in
the analysis of future LBL experiments is the presence of parameter
degeneracies, which pose a serious limitation to the determination of
$\theta_{13}$, $\dcp$ and the sign of $\Dmq_{31}$.
This problem is illustrated in Fig.~\ref{fig:degeneracy}, where we
show the allowed regions in the plane of $\sin^22\theta_{13}$ and
$\dcp$ for $\beta$B, SPL and T2HK.
As visible in this figure, for the $\beta$B the intrinsic degeneracy
cannot be resolved, while for the super beam experiments SPL and T2HK
there is only a four-fold degeneracy related to the sign of
$\Dmq_{31}$ and the octant of $\theta_{23}$. Once atmospheric data are
included in the fit all the degeneracies are nearly completely
resolved, and the true solution is identified at 95\%~CL. This clearly
show the presence of a synergy between atmospheric and long-baseline
data: at least for this specific example, the combination of the two
sets is much more powerful than the simple sum of each individual data
sample.

\section{Conclusions}
\label{sec:conclusions}

In this talk we have discussed the potentialities of atmospheric
neutrino data in the context of future neutrino experiments. We have
shown that despite the large uncertainties in the neutrino fluxes
atmospheric data will still provide useful information on the neutrino
parameters, due to their very broad range in neutrino energy and nadir
angle. In particular, we have proved that the sensitivity obtained by
a combination of atmospheric and long-baseline data is much stronger
than the one achievable by each data set separately.

\medskip

Work supported by MICINN through the Ram\'on y Cajal program and through
the national project FPA2006-01105, and by the Comunidad Aut\'onoma de
Madrid through the HEPHACOS project P-ESP-00346.

\end{document}